\documentclass[%
aps,
prl,
reprint,
preprintnumbers,
nofootinbib,
 amsmath,amssymb,
]{revtex4-2}

\usepackage{graphicx}
\usepackage{dcolumn}
\usepackage{bm}
\usepackage{slashed} 
\usepackage{hyperref}
\hypersetup{pdfstartview=FitV,colorlinks=true,linkcolor=blue,citecolor=red,filecolor=black,urlcolor=blue}
\usepackage[dvipsnames]{xcolor}
\usepackage{subfig}

\begin{document}

\title{
Post-Reheating Inflaton Production as a Probe of Reheating Dynamics
}

\author{Kunio Kaneta}
    \affiliation{Faculty of Education, Niigata University, Niigata 950-2181, Japan}
\author{Tomo Takahashi}
    \affiliation{Department of Physics, Saga University, Saga 840-8502, Japan}
\author{Natsumi Watanabe}
    \affiliation{Graduate School of Science and Engineering, Saga University, Saga 840-8502, Japan}

\date{\today}

\begin{abstract}
    Cosmological reheating bridges the inflationary epoch and the hot big bang phase, yet its underlying dynamics remain poorly understood. In this work, we investigate a minimal scenario in which the inflaton evolves under a simple power-law potential during reheating and interacts with other particles via renormalizable couplings. We show that inflaton quanta can be regenerated from the thermal bath even after the decay of the coherent inflaton field, unveiling a previously overlooked channel for inflaton particle production, which offers a novel window into probing reheating via consistency with observations and laboratory experiments.
    Remarkably, this mechanism may also account for the observed dark matter abundance, providing a natural link between early Universe dynamics and present-day cosmological observations.
\end{abstract}

\maketitle

{\it Introduction.---}
The hot big bang Universe has been at a central place in modern cosmology since the success of the Big Bang Nucleosynthesis~\cite{Alpher:1948ve,Hayashi:1950lqo,Alpher:1950zz} and subsequent observations of the cosmic microwave background (CMB) radiation~\cite{Penzias1965}.
The cosmic inflation has been proposed to keep the success of and complete the hot big bang scenario by solving the horizon and flatness problems~\cite{Guth:1980zm,Sato:1981qmu,Linde:1981mu,Linde:1983gd}, while seeding the primordial density perturbations that lead to the large-scale structure of the Universe~\cite{Bardeen:1980kt,Mukhanov:1981xt,Starobinsky:1982ee,Guth:1982ec,Linde:1982uu}.
Once inflationary expansion ends, the Universe is left cold and empty, and thus a heating mechanism, called reheating, is necessary to bring the Universe into a hot state.

The reheating era bridges inflation and the radiation-dominated Universe, yet its microphysics remains elusive. In conventional scenarios, the inflaton condensate decays completely~\cite{Dolgov:1982th,Abbott:1982hn,Nanopoulos:1983up}, producing the Standard Model particles and reheating the Universe to a temperature $T_{\rm reh}$.
However, this picture may be incomplete.
There may be preheating regime prior to reheating, where the inflaton field oscillates about its minimum and decays into other particles through non-perturbative processes~\cite{Kofman:1994rk,Kofman:1997yn,Traschen:1990sw,Yoshimura:1995gc}.
Such a non-perturbative process may lead to parametrically enhanced superradiance of inflaton as well~\cite{Yoshimura:2024rgl}.
There may also be a possibility that the inflaton condensate fragments into inflaton particles before reheating completes~\cite{Amin:2010xe,Amin:2010dc,Lozanov:2017hjm,Garcia:2023dyf}.
While the inflaton potential fixes the reheating dynamics in principle, modeling preheating, fragmentation, and thermalization remains highly uncertain. We therefore assume efficient reheating and focus on the novel aspect of our work: inflaton regeneration after reheating as a new window probing reheating model and its implications for dark matter and collider phenomenology.
Given the observational challenges associated with probing the reheating epoch, any potential signature from this phase is of significant interest.

In this {\it letter}, we propose that the inflaton quanta can be regenerated from thermal bath {\it in the post-reheating era} via the same coupling responsible for rehearing, provided the effective mass of the inflaton is smaller than the bath temperature.
This condition is commonly realized when the inflaton potential takes a monomial form higher than a quadratic term: $V (\phi) \propto \phi^k$  with $k = 4, 6, \cdots$ near the bottom of the potential. 
The CMB does not constrain the shape of the potential around its minimum, and a pure monomial extended to large field values is excluded by the current limits on the tensor-to-scalar ratio.
What the CMB does constrain is $k$ within families of models in which a single exponent controls both the inflationary plateau and the region around the minimum, such as the T-model of Eq.~(\ref{eq:inflaton potential}) adopted below.
For that family, the recent measurements of the spectral index $n_s$~\cite{AtacamaCosmologyTelescope:2025nti,AtacamaCosmologyTelescope:2025blo,SPT-3G:2025bzu,Balkenhol:2025wms} favor $k\ge6$ at
the 68\% CL, with $k=4$ still allowed at the 95\% CL~\cite{Kaneta:2026qxx}.
The abundance of produced inflaton quanta can be non-negligible, even leading to overclose the Universe, depending on the size of the coupling for a relevant interaction, from which we may place constrains on the interactions responsible for reheating. This opens up a new avenue for probing reheating processes and offers a valuable guidance for model-building of reheating scenarios motivated by particle physics.

\bigskip
{\it Evolution of inflaton condensate.---}
We introduce the inflaton condensate as a zero-mode of $\phi(x)$:
\begin{align}
   \phi(x) &= \overline\phi(t) + \delta\phi(x),
\end{align}
where $\overline\phi$ and $\delta\phi$ denote the zero mode and the non-zero momentum modes (or inflaton quanta), respectively.
Note that since $\delta\phi$ does not have the zero-momentum mode, $\overline\phi$ and $\delta \phi$ are orthogonal to each other.

The Lagrangian of $\overline\phi$ relevant for inflation is given by
\begin{align}
   {\cal L}_{\rm inf}
   &=
   \frac{1}{2}\dot{\overline\phi}^2 - V(\overline\phi),
\end{align}
where, during reheating ($\overline\phi\ll M_P$), we consider a simple power-law potential
\begin{align}
\label{eq:V}
   V(\overline\phi) &= \lambda M_P^4 \left(
      \overline\phi/M_P
   \right)^k
\end{align}
with $k$ an even integer and $M_P$ the reduced Planck mass, $M_P\simeq 2.4\times10^{18}$ GeV.
From the agnostic viewpoint, inflaton potential during reheating given above is not necessarily related to the inflationary predictions. Nonetheless, when the above term is somehow relevant during inflation, the inflaton potential coupling $\lambda$ is determined by CMB observation.
For definiteness, we suppose that the inflaton potential during inflation ($\overline\phi \gg M_P$) is given by an extended version of the T-model~\cite{Kallosh:2013hoa},
\begin{align}
   V(\overline\phi) &= \lambda M_P^4 \left[
      \sqrt{6}\tanh\left(
         \frac{\overline\phi}{\sqrt{6}M_P}
      \right)
   \right]^k.
   \label{eq:inflaton potential}
\end{align}
In this case, $\lambda$ is approximately estimated by
\begin{align}
   \lambda &\simeq \frac{18\pi^2 A_S}{6^{k/2}N^2},
\end{align}
where the amplitude of scalar perturbation is given by $A_S\simeq 2.1\times 10^{-9}$ at the Planck pivot scale $k_{\rm pivot}=0.05~{\rm Mpc}^{-1}$~\cite{Planck:2018jri}.
The number of $e$-folds between horizon exit of the pivot scale and the end of inflation, $N$, depends on models of inflation and reheating dynamics.
For $N=50$ as an example\footnote{
Note that precise determination of $N$ refines the inflationary predictions, namely, $\lambda$ in Eq. (\ref{eq:inflaton potential}), whose contribution, nevertheless, does not affect the following argument considerably.
}, we obtain $\lambda = 4.1\times10^{-12}, 6.9\times 10^{-13}, 1.2\times 10^{-13}, 1.9\times 10^{-14}$ for $k=4, 6, 8, 10$, respectively.

After inflation ends at the scale factor $a=a_e$, the energy density of $\overline\phi$ evolves as
\begin{align}
   \rho_{\overline\phi}(a)
   &=
   \rho_e \left(
      \frac{a}{a_e}
   \right)^{-\frac{6k}{k+2}},
\end{align}
where $\rho_e\equiv \rho_{\overline\phi}(a_e)$.
As studied in detail in Refs.~\cite{Ichikawa:2008ne,Garcia:2020eof,Garcia:2020wiy}, while its evolution resembles dust-like energy density for $k=2$, it redshifts away faster for $k\geq 4$.
Remarkably, whereas the inflaton mass is constant for $k=2$, it decreases with time when $k\geq 4$ and becomes negligible in later times.
As we will argue, because the negligible inflaton mass is crucial in regenerating the inflaton quanta, we focus on $k\geq 4$ in the rest of our analysis.

\bigskip
{\it Inflaton interactions.---}
Inflaton must have some coupling(s) through which radiation-dominated Universe is eventually realized such that the Universe can be described by the standard big bang cosmology.
Although the process of reheating is generally model-dependent, if we limit ourselves to the renormalizable interactions, we can consider only the following type: $y\phi\overline f f$, $\mu\phi \chi \chi$, $\sigma \phi^2 \chi^2$, where $\chi$ and $f$ are bosonic and fermionic field constituting the thermal plasma, respectively and $y,\mu$ and $\sigma$ are the coupling parameters.  In the following, we assume that the reheating is somehow completed via one of the interactions above at $a=a_{\rm reh}$. 

As mentioned above, we consider the potential~\eqref{eq:V} with $k \ge 4$,which leads to a time-dependent inflaton mass that decreases as the Universe expands:
\begin{align}
   m_\phi(a) &= m_\phi(a_e)\left(
      \frac{a}{a_e}
   \right)^{-3\frac{k-2}{k+2}} \,.
\end{align}
Note that $a_e$ is defined by the end of the slow-roll regime, specifically when the slow-roll parameter $\epsilon(\phi)$ satisfies $\epsilon(\phi)= 2M_P^2(\partial_\phi H)^2/H^2=1$ with $H$ being the Hubble parameter.
Although $a_e$ serves as a reference point, physical quantities such as the inflaton mass at $a_e$ are determined by the field value $\overline\phi_e\equiv \overline\phi(a_e)$.
For instance, the inflaton mass at $a=a_e$ is given by $m_\phi(a_e)=\sqrt{\lambda k(k-1)(\overline\phi_e/M_P)^{k-2}}M_P= 2.5\times10^{13}~{\rm GeV}, 4.2\times10^{13}~{\rm GeV}, 7.5\times10^{13}~{\rm GeV}, 1.4\times10^{14}~{\rm GeV}$ for $k=4,6,8,10$, respectively.

The coherent oscillation of $\overline\phi(t)$ ceases when reheating is completed, and thus the inflaton can be effectively massless.
Our scenario assumes that $\chi$ develops a vacuum expectation value (VEV) during the radiation-dominated era to generate the mass for $\delta\phi$ if $\phi^2\chi^2$ coupling is present\footnote{
Note that we do not assume a bare mass term for inflaton. Its effect on reheating has been studied in Ref.~\cite{Clery:2024dlk}.
}.
This indicates that the same interaction relevant for the reheating regenerates inflaton quanta during {\it post-reheating era}, and its abundance can become significant to affect the evolution of the Universe, from which we can constrain the coupling responsible for reheating, and probe the reheating process. 

Here we consider the interaction $-{\cal L}_{\rm int} = (1/2) \sigma \phi^2 \chi^2$ with $\sigma$ being a coupling constant as an explicit example to demonstrate how post-reheating regeneration of the inflaton quanta allows us to study the reheating process. 
Since this interaction has $Z_2$ symmetry\footnote{In general, this symmetry may not be respected in the inflaton potential.}, making it stable (see e.g., \cite{Mukaida:2013xxa} for some particular mechanism), $\delta\phi$ may become dark matter or dark radiation, depending on the mass and VEV of $\chi$. The abundance can be calculated by the standard procedure using the Boltzmann equation, in which  the $\delta\phi$ number density $n_{\delta\phi}$ is obtained by solving~(e.g., \cite{Bernstein:1988bw}):
\begin{align}
   \frac{dn_{\delta\phi}}{dt}+3Hn_{\delta\phi}
   &=
\gamma_{\delta\phi\delta\phi\leftrightarrow\chi}+
\gamma_{\delta\phi\delta\phi\leftrightarrow\chi\chi}
\end{align}
where the right-hand side includes the contributions from the processes $\delta\phi\delta\phi\leftrightarrow \chi$ and $\delta\phi\delta\phi\leftrightarrow\chi\chi$ whose reaction rates are denoted by $\gamma_{\delta\phi\delta\phi\leftrightarrow\chi}$ and $\gamma_{\delta\phi\delta\phi\leftrightarrow\chi\chi}$ respectively, and $H$ is the Hubble parameter.
If $\chi$ is identified as the Standard Model Higgs boson, additional channels of $\delta\phi\delta\phi\leftrightarrow$ (a pair of the Standard Model particles) also contribute.
The details of the analysis are presented in Ref.~\cite{Kaneta:2026qxx}.
Note that the $\chi\leftrightarrow \delta\phi\delta\phi$ process is induced by the VEV of $\chi$.
Using the thermally-averaged production rate $\langle\sigma_{\delta\phi\delta\phi}v_{\rm rel}\rangle$, the Boltzmann equation can be expressed as
\begin{align}
   \frac{d Y}{dT} &= -\frac{\langle\sigma_{\delta\phi\delta\phi} v_{\rm rel}\rangle s}{H T}\left(
      Y_{\rm eq}^2-Y^2
   \right),
   \label{eq: Boltzmann eq}
\end{align}
where $Y=n_{\delta\phi}/s$ with the entropy density $s$ and $Y^{\rm eq}$ being its equilibrium value.
By solving Eq.~(\ref{eq: Boltzmann eq}) under the assumption that $\chi$ is in thermal equilibrium and develops a VEV $v$ at some epoch, thereby generating the mass for the inflaton as $m_\phi=\sqrt{\sigma}v$, we evaluate the final abundance of $\delta \phi$ given by 
\begin{align}
   \Omega_{\delta\phi}h^2 &= \frac{m_{\phi}n_{\delta\phi}}{\rho_{c0}{h^{-2}}} = 2.89\times 10^{8} \frac{m_{\phi}}{\rm GeV}Y_0,
\end{align}
where $\rho_{c0}$ is the today's critical density, and we define $Y_0=Y(T_0)$ with $T_0$ the temperature at the present Universe.

\begin{figure}[t]
   \centering
   \includegraphics[width=.45\textwidth]{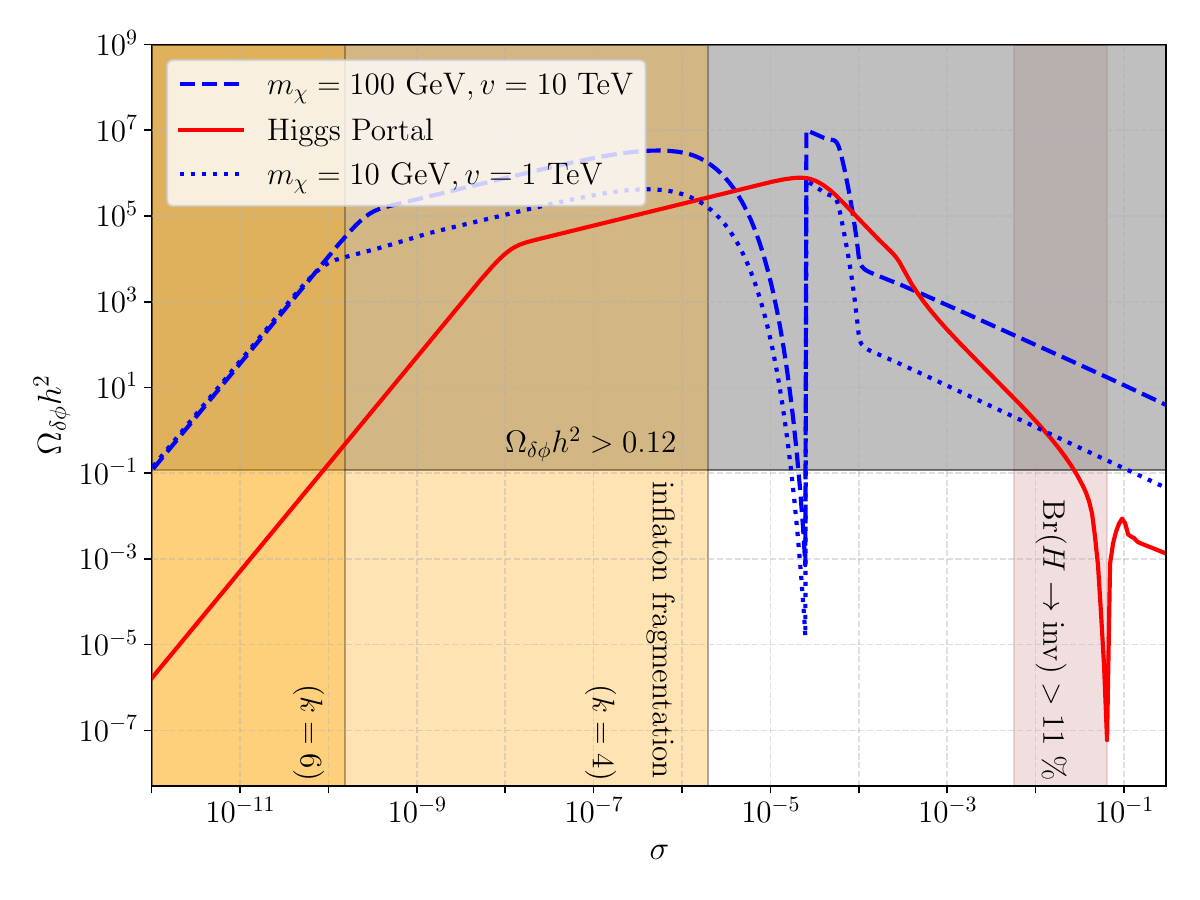}
   \caption{Inflaton abundance for different values of $v$. The constraint from the Higgs invisible decay may apply for only the case where $\chi$ is identified as the Standard Model Higgs boson.}
   \label{fig: Oh2 for sigma}
\end{figure}

Figure~\ref{fig: Oh2 for sigma} shows the $\delta\phi$ abundance with $\chi$ being a generic scalar depicted by the blue dashed line ($m_\chi=100~{\rm GeV}$ and $v=10~{\rm TeV}$) and the blue dotted line ($m_\chi=10~{\rm GeV}$ and $v=1~{\rm TeV}$), as well as the case where $\chi$ is identified as the Standard Model Higgs boson depicted by the red line ($m_\chi=m_h=125~{\rm GeV}$ and $v=246~{\rm GeV}$).
See Ref.~\cite{Kaneta:2026qxx} for the details, including analytic estimate of the abundance, and the results for other parameter space.
For $\sigma\gtrsim 10^{-6}$, $\delta\phi$ reaches thermal equilibrium, and thus the abundance is determined by ``freeze-out" mechanism\footnote{Scenarios of inflaton as a freeze-out dark matter have been studied in Refs.~\cite{Lerner:2009xg,Okada:2010jd,Mukaida:2014kpa}}, which may be regarded as a special case of WIMPflation~\cite{Hooper:2018buz,Garcia:2021gsy}. 
Note that the sharp drop of the abundance is due to the resonance of the $\chi$ particle, which happens when $m_\phi\simeq m_\chi/2$.
In contrast, in the ``freeze-in" regime, $\delta\phi$ is never thermalized.
The gray region, where $\Omega_{\delta\phi}h^2 > 0.12$~\cite{Planck:2018vyg}, is excluded\footnote{Our analysis assumes no entropy injection after the inflaton decoupling. Note that additional entropy dilution may open additional parameter space.}, yielding an upper bound on $\sigma$. 
For simplicity, we assume a vanishing initial inflaton density at the onset of reheating. This assumption may be altered by non-perturbative preheating and fragmentation~\cite{Amin:2010dc,Amin:2010xe,Lozanov:2017hjm,Garcia:2023dyf}, which can produce inflaton particles efficiently. 
Since our aim is to determine the abundance itself rather than to bound it, what matters is the size of the population that survives until the freeze-in takes place.
For the potential of Eq.~\eqref{eq:V}, the self-resonance driven by the rapidly varying inflaton mass is not a channel separate from fragmentation, since lattice studies show that the broad resonance band shrinks as the Universe expands and that it is the first narrow band which slowly drives the condensate to backreaction and fragmentation~\cite{Lozanov:2017hjm}.
Whatever its origin, the population of inflaton quanta left at the end of reheating enters Eq.~\eqref{eq: Boltzmann eq} as an initial condition $Y_{\rm ini}$, which is then depleted by the field dependent inflaton mass.
Extrapolating the depletion computed in Ref.~\cite{Garcia:2023dyf}, Ref.~\cite{Kaneta:2026qxx} finds the surviving abundance to be negligibly small for $k\ge 6$, and thus the final abundance is determined by the post-reheating regeneration process.
This does not require reheating to be completed before the condensate fragments, since what enters the final abundance is the population that survives the depletion rather than the amount initially produced.

Since we assume the potential given in Eq.~\eqref{eq:V} with $k\ge 4$,  i.e., having the inflaton self-interaction, the homogeneous inflaton field $\overline\phi$ can eventually fragment, depending on the coupling $\sigma$. If reheating has not yet completed before fragmentation occurs, the resulting inflaton quanta $\delta\phi$ dominate the total energy density of the Universe~\cite{Garcia:2023dyf}.
To ensure that reheating completes before fragmentation, the coupling $\sigma$ needs to be greater than certain values, as discussed in Ref.~\cite{Garcia:2023dyf}.
Notice that the inflaton dissipation rate during reheating is characterized by $\sigma_{\rm eff}$, which accounts for the time-averaged inflaton oscillations and is related to the bare coupling $\sigma$~\cite{Garcia:2020eof,Garcia:2020wiy}.
In Ref.~\cite{Garcia:2023dyf}, bounds on the effective coupling $\sigma_{\rm eff}$ is provided, which can be translated to bounds on the bare coupling as $\sigma_{\rm eff}/\sigma=3.6, 7.0, 11, 15$ for $k=4, 6, 8, 10$, respectively. 
These bounds are illustrated in Fig.~\ref{fig: Oh2 for sigma} for the cases with $k =4$ and $6$, which correspond to $\sigma<2.0\times10^{-6}$ and $\sigma<1.5\times10^{-10}$, respectively. 
From the figure, we also observe that, except for the resonant regime, the boundary of the gray region marks the couplings at which $\delta\phi$ accounts for the entire dark matter abundance.
For $(m_\chi,v) = (10~{\rm GeV}, 1~{\rm TeV})$ and $(125~{\rm GeV}, 246~{\rm GeV})$ this occurs at $\sigma\simeq0.1$ and $2.7\times10^{-2}$, respectively, where the abundance is set by freeze-out, and, in the latter case, also at $\sigma\simeq10^{-10}$, where it is set by the post-reheating regeneration.
For $k\ge6$ the initial abundance left by fragmentation is negligible, as discussed above, and both of these solutions are available.
For $k=4$ the fragmentation threshold, $\sigma=2.0\times10^{-6}$, lies above the entire freeze-in regime, where the initial abundance is then not guaranteed to be negligible, and only the freeze-out solution remains.
Importantly, the reheating process is therefore relevant to the later evolution of the Universe through the regeneration of inflaton quanta.

Although the above general consideration has already provided substantial insight into the reheating process, we can obtain more constraints on the reheating coupling by specifying the model further. Given that the Standard Model particles should be eventually produced, one of the simplest scenarios would be to identify $\chi$ as the Standard Model Higgs boson, which we are going to investigate in the following.

\bigskip
{\it Higgs portal interactions.---}
The idea of Higgs portal interactions to particles beyond the Standard Model has been widely studied in the context of dark matter~\cite{Silveira:1985rk,McDonald:1993ex,Burgess:2000yq,Djouadi:2011aa}, and in the context of inflation, the effect of the Higgs portal couplings to inflationary models and subsequent reheating has been studied in Refs.~\cite{Lebedev:2011aq,Mukaida:2014kpa,Drees:2021wgd,Lebedev:2021zdh,Yang:2023rnh,Cado:2025orb}. Here we identify $\chi$ as the Higgs boson, being responsible for reheating. At the renormalizable level, $\phi$ cannot have a coupling to the Standard Model fermions due to gauge symmetry, and thus we only consider the bosonic coupling below, which we label by b-I and b-II:
\begin{align}
   -{\cal L}_{\rm int} &=
   \begin{cases}
      \sigma \phi^2 |H|^2 & \text{(b-I)} \\
      2\mu \phi |H|^2 & \text{(b-II)}
   \end{cases},
\end{align}
where $H$ is the Higgs doublet. Now we discuss what information can be obtained for each case in the following.

{\it Case b-I.---}
This case is essentially the same as the one discussed above for the interaction $-{\cal L}_{\rm int} = (1/2) \sigma \phi^2 \chi^2$, but with the VEV $v=246~{\rm GeV}$. Thus the abundance of $\delta \phi$ produced from this interaction is given by red line in Fig.~\ref{fig: Oh2 for sigma}. Since we identify $\chi$ as the Higgs boson, collider search can also bring us more insight into the coupling. Indeed the Higgs invisible decay measurement excludes the parameter
$5.6\times10^{-3}<\sigma<6.5\times10^{-2}$ by requiring the branching ratio to be less than 11
~\%~\cite{ATLAS:2023tkt,Arcadi:2019lka}
, which is also shown in Fig.~\ref{fig: Oh2 for sigma}. It should be noted that, in the region where $m_{\phi}\simeq m_h/2$, the Higgs resonance may enhance the reaction rate and suppress the abundance considerably, which happens when $\sigma\simeq 0.065$.  However, this region has already been excluded by the Higgs invisible decay measurement mentioned above, and thus it is irrelevant to our discussion. We also note that the other $s$-channel processes of pair-annihilating $\delta\phi$ into a pair of the Standard Model particles can also be included, however, its rate is suppressed by $m_{\phi}^2/m_h^2$ (in non-relativistic regime) for $m_{\phi}\ll m_h$~\cite{Burgess:2000yq} and hence it is neglected in our analysis.

{\it Case b-II.---}
A tadpole term develops after the Higgs field acquires a non-zero VEV, yielding a non-zero VEV of $\phi$,
\begin{align}
   \langle\phi\rangle &= \left(
      \frac{-v^2\mu}{k\lambda M_P^3}
   \right)^{\frac{1}{k-1}}M_P,
\end{align}
which induces a mass of $\delta\phi$ given by
\begin{align}
   m_\phi^2 &= k(k-1)\lambda M_P^2\left(
      \frac{-\mu v^2}{k\lambda M_P^3}
   \right)^{\frac{k-2}{k-1}}.
\end{align}
However, such a contribution is negligibly small in the parameter space of our interest, compared to the radiatively induced mass as discussed in the following.
Therefore, we drop the tadpole contribution to $\langle\phi\rangle$ in the following analysis.

The Coleman-Weinberg potential of $\phi$, generated by generic $\chi$ fields at one loop level, is given by
\begin{align}
   V_{\rm CW}(\phi) &= c_\chi \frac{(2\mu\phi)^2}{64\pi^2}\left(
      \ln\frac{2\mu\phi}{\Lambda^2}-\frac{3}{2}
   \right),
\end{align}
where $\Lambda$ is the renormalization scale and $c_\chi$ is the number of degrees of freedom. For the Higgs boson case, we take $c_\chi=4$.
By minimizing $V+V_{\rm CW}$, we find the effective mass of inflaton as
\begin{align}
   m_\phi^2 &= \frac{\mu^2}{\pi^2}(1 + \cdots),
\end{align}
where the dots include $\Lambda$-dependent terms in the form of Lambert $W$ function.
Given that $\Lambda$ may take at an arbitrary scale, we may set such terms smaller than unity, so that the effective mass is simply given by
\begin{align}
   m_\phi &\simeq \frac{\mu}{\pi}.
\end{align}

The physical Higgs field $h$ mixes with the inflaton particle $\delta\phi$ through
\begin{align}
   {\cal L}_{\rm mix} &= \frac{1}{2}(2\mu v) \delta\phi h,
\end{align}
which induces a mixing angle
\begin{align}
   \theta \simeq \frac{2\mu v}{m_h^2},
\end{align}
under the assumption $m_\phi\sim\mu\ll m_h$, with the Higgs mass $m_h=125$ GeV.
The inflaton mass is set by $\mu$ in the present case, and hence we obtain $\theta\simeq m_\phi/10~{\rm GeV}$.
Restricting to $m_\phi > 50$ MeV then gives $\theta\gtrsim 5\times10^{-3}$, which is excluded by meson-decay measurements~\cite{Ferber:2023iso}.
For smaller masses and mixing, stringent astrophysical and cosmological bounds apply~\cite{Beacham:2019nyx} (see also~\cite{Ibe:2021fed} for further details).

These constraints can be relaxed if the inflaton mass is not tied to $\mu$.
A simple possibility is to introduce a $\sigma \phi^2|H|^2$ interaction in addition to the $\mu \phi |H|^2$ term, which generates a $\mu$-independent contribution to $m_\phi$. 
In such a setup, the $\sigma$ coupling can control inflaton production while the $\mu$ coupling governs inflaton decay, potentially leading to richer phenomenology. 
Alternatively, one may simply add a bare mass term, $m_\phi^2\phi^2/2$.
In any case, treating $\theta$ and $m_\phi$ as independent parameters, we obtain a constraint on $\mu$ as
\begin{align}
   \mu \lesssim 10^{-3}\frac{m_h^2}{2v} \simeq 0.032~{\rm GeV}.
\end{align}
from the dark Higgs search~\cite{Ferber:2023iso}.

The limit from the inflaton fragmentation is also provided for the effective coupling $\mu_{\rm eff}$ which corresponds to the bare coupling $\mu$ as $\mu_{\rm eff}/\mu=1.4, 1.7, 1.9, 2.1$ for $k=4,6,8, 10$ \cite{Garcia:2020wiy,Garcia:2023dyf}.
The lower bound on $\mu_{\rm eff}$ for $k=4,6$ is well above the EW scale \cite{Garcia:2023dyf}, hence our analysis is not relevant for these limits.
On the other hand, for $k = 8, 10$, the lower bound is given by $\mu = 0.33~{\rm GeV}, 7.1\times10^{-7}~{\rm GeV}$, respectively.
Notice that the case with $k=8$ is already disfavored by the dark Higgs search.

\begin{figure}[t]
   \centering
   \includegraphics[width=.45\textwidth]{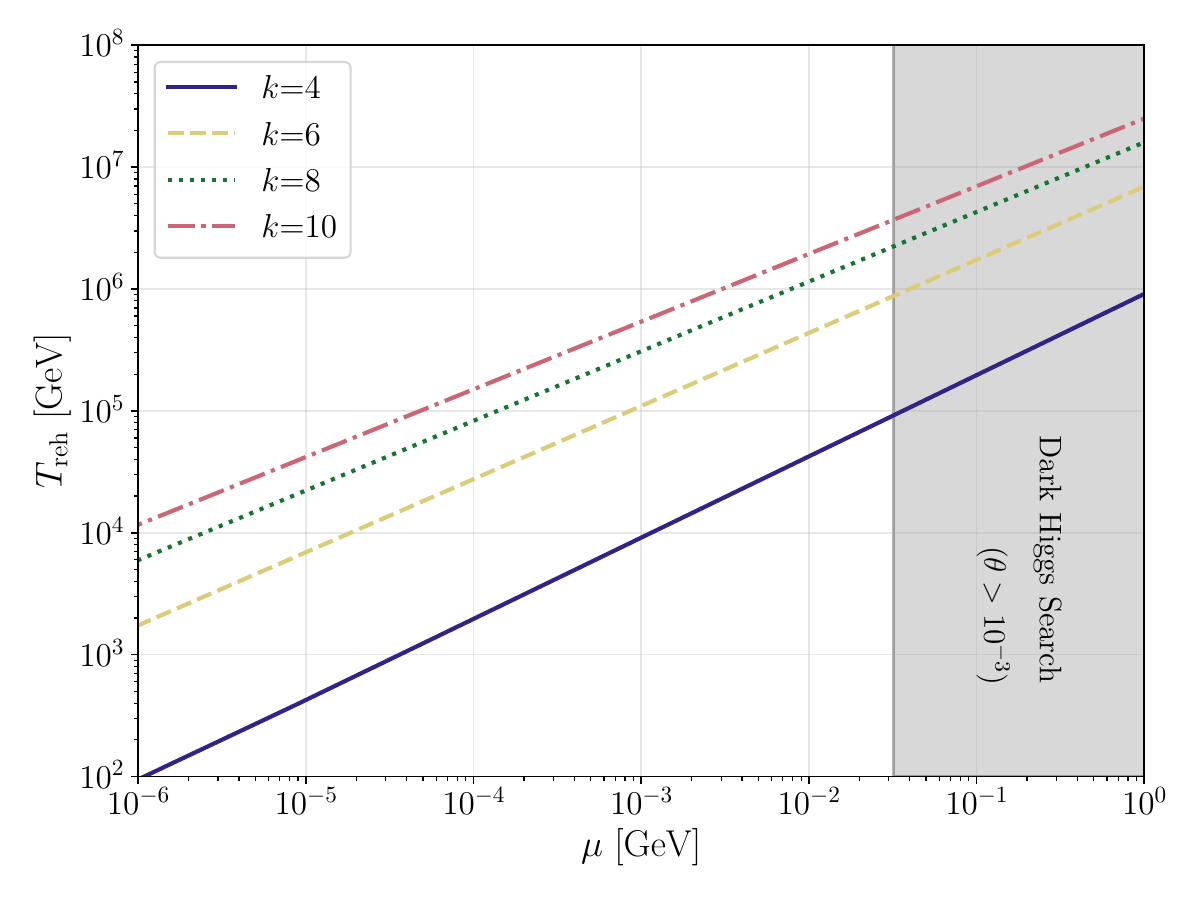}
   \caption{Reheating temperature and constraint from the dark Higgs search.}
   \label{fig: T_reh and dark Higgs search constraint}
\end{figure}

If we neglect the fragmentation, we can perform numerical estimation of the reheating temperature as a function of $\mu$, following the approach in Ref.~\cite{Garcia:2020wiy}.
This approximation is justified since fragmentation would have only a minor impact on reheating for this type of coupling, as discussed in Ref.~\cite{Garcia:2023dyf}.
Note that the lines of $k=4,6,8,10$ in Fig.~\ref{fig: T_reh and dark Higgs search constraint} are obtained by solving the consistency relation among $N$, $T_{\rm reh}$ and the equation-of-state parameter during reheating, $w_{\rm reh}$, which is determined by $k$ as $w_{\rm reh}=(k-2)/(k+2)$~\cite{Garcia:2020eof,Garcia:2020wiy}.
See Ref.~\cite{Kaneta:2026qxx} for the details.
Figure~\ref{fig: T_reh and dark Higgs search constraint} shows the constraint from the dark Higgs search on the reheating temperature, which places an upper bound as
$T_{\rm reh}<9.2\times10^{4}~{\rm GeV}, 8.8\times10^{5}~{\rm GeV}, 2.2\times10^{6}~{\rm GeV}, 3.7\times10^6~{\rm GeV}$ for $k=4, 6, 8, 10$, respectively.
Note again that our analysis may apply for only the case of $m_\phi\ll m_h$.

\bigskip
{\it Conclusion.---}
We have shown that a significant regeneration of inflaton quanta during post-reheating era can serve as a powerful probe into the reheating process. Such a regeneration of inflaton quanta typically occurs when the (effective) mass of the inflaton is smaller than the temperature of the thermal bath, as is often the case in inflation models where a self-interaction term dominates near the potential minimum. By examining specific interaction examples, we demonstrated how such regeneration can be used to constrain couplings relevant to reheating.

Furthermore, we have investigated a scenario in which the Standard Model Higgs boson is coupled to the inflaton, enabling reheating via the Higgs portal. In this framework, collider constraints, such as the Higgs invisible decay and the dark Higgs search, can be combined with the consideration of the inflaton quanta regeneration to further refine our understanding of reheating dynamics.

\bigskip
{\it Acknowledgments.---}
We would like to thank Yann Mambrini and Michael Schmidt for useful discussion.
This work was made possible by Institut Pascal at Universit\'e Paris-Saclay with the support of the program “Investissements d’avenir” ANR-11-IDEX-0003-01, the P2I axis of the Graduate School of Physics of Universit\'e Paris-Saclay, as well as IJCLab, CEA, IAS, OSUPS, and APPEC. This work was also supported by JSPS KAKENHI Grant Numbers 25K01004 (TT) and MEXT KAKENHI 23H04515 (TT), 25H01543 (TT).
KK would like to express special thanks to the Mainz Institute for Theoretical Physics (MITP) of the Cluster of Ecellence PRISMA$^+$ (Project ID 390831469), for its hospitality and support.

\bibliography{biblio_letter}

\end{document}